\documentclass{elsart}
 
\usepackage[T1]{fontenc}
\usepackage{amsmath,amssymb}
\usepackage[numbers,sort]{natbib}

\newcommand*{\vect}[1]{\mathbf{#1}}      
\newcommand*{\abs}[1]{\lvert{#1}\rvert}  
\newcommand*{\bra}[1]{\langle{#1}\rvert}
\newcommand*{\ket}[1]{\lvert{#1}\rangle}

\begin{document}

\bibliographystyle{my-hunsrt}

\begin{frontmatter}

\title{Color van der {W}aals forces between heavy quarkonia in
  effective {QCD}}
\author{Jakub Nar\k{e}bski}
\address{
  Institute of Theoretical Physics, Warsaw University,
  ul. Ho\.{z}a 69, 00-681 Warsaw, Poland}
\ead{jnareb@fuw.edu.pl}

\date{25 September 2002}

\begin{abstract}
  The perturbative renormalization group for light--front QCD
  Hamiltonian produces a logarithmically rising interquark potential
  already in second order, when all gluons are neglected.  There is a
  question if this approach produces also color van der Waals forces
  between heavy quarkonia and of what kind.  This article shows that
  such forces do exist and estimates their strength, with the result
  that they are on the border of exclusion in naiv\'{e} approach,
  while more advanced calculation is possible in QCD.
\end{abstract}



\maketitle

\end{frontmatter}

In the constituent models used to describe hadrons, quarks can be
assumed to interact via additive 2--body confining potential with the
same color structure as if the interaction were mediated by one--gluon
exchange,
\begin{equation}
  \label{eq:potential-model}
  H_I = - \sum_{i<j} F_i \cdot F_j \; V(\vect{r}_i - \vect{r}_j),
\end{equation}
where $F_i \cdot F_j$ denotes $\sum_{a=1}^{8} F^a_i F^a_j$, and
$F^a_i$ are the eight SU(3) color generators for $i$--th quark or
antiquark, located at $\vect{r}_i$.  The potential models describe
properties of single--hadron states quite well, but also predict
power--law color van der Waals forces between colorless hadrons, which
contradict experiments by being too strong.  The color confining
potentials also cause that the model Hamiltonians are unbounded from
below for all states except color singlets, triplets and
antitriplets~\citep{Greenberg:1981xn}.  Potential models which try to
avoid these drawbacks appear to be arbitrary and not systematically
related to QCD.

One of the attempts to derive a constituent picture of hadrons
directly from QCD is by using the perturbative similarity
renormalization group concept for light--front
Hamiltonian~\citep{Wilson:1994fk}.  \citet{Perry:1997uv}, and
\citet{Brisudova:1996hv} have shown that one can get confining
potential already in second--order calculations based on this concept.
In their calculation the model Hamiltonian is bounded from below for
all color states of quark--antiquark pairs.  However, their approach
was not boost invariant and it was not clear how to describe
two--meson states with clear control on their masses.  Since then, the
similarity renormalization group approach has advanced so that one can
preserve boost invariance and required cluster properties by
introducing effective particles~\citep{Glazek:1998sd,Glazek:2001gb}.

We use similarity renormalization approach (see
Ref.~\citep{Glazek:2001gb} for description and references) to express
the field--theoretical canonical Hamiltonian in terms of creation and
annihilation operators for effective particles, which are unitarily
equivalent to the canonical ones, viz.\ 
\begin{math}
  b_{\lambda} = 
  U^{\dag}_{\lambda} b_{\text{can}} U^{\hphantom{\dag}}_{\lambda}
\end{math}
etc.  The Hamiltonian $H_\lambda$ written using the new operators is
band--diagonal, i.e.\ each term is multiplied by formfactor
$f_{\lambda}$, which vanishes when energy changes by more than width
$\lambda$.  We construct similarity transformation
$U^{\vphantom{\dag}}_{\lambda}$ using second order perturbation
theory.

Following~\citep{Wilson:1994fk}, we start from the canonical QCD
Hamiltonian on the light front (LF), taking the advantage of the fact
that in this formulation vacuum can be made trivial by means of a
cutoff on longitudinal momenta.  We regulate the Hamiltonian term by
term in the expansion into products of creation and annihilation
operators~\citep{Glazek:2001gb}, introducing regulating factors
\begin{math}
  r_{\Delta\delta} (\kappa^\perp, x) = 
  \exp\bigl[\kappa^{\perp\,2}/(x\,\Delta^2)\bigr]
  r_{\delta}(x)
\end{math}
for each operator in every term.  Here $x$ and $\kappa^{\perp}$ are
the relative momenta of a particle with respect to other particles in
a vertex (see~\citep{Glazek:2001gb}).  In gauge--field theory we need
$r_{\delta}$, the small--$x$ regulator, and we choose 
$r_{\delta}(x) = \Theta(x - \delta)$ as the simplest option.  We use
boost--invariant  formfactor~\citep{Glazek:2001gb}
\begin{equation}
  \label{eq:f_lambda}
  f_{\lambda}(\mathcal{M}^2_c - \mathcal{M}^2_a) = 
  \exp \bigl[-(\mathcal{M}^2_c - \mathcal{M}^2_a)^2 \big/ \lambda^4\bigr].
\end{equation}
Here $\mathcal{M}^2_c \equiv \bigl(\sum_{i\in\text{creat}} p_j \bigr)^2$ 
is the free invariant mass of the particles created in the vertex, 
and $\mathcal{M}^2_a$, similarly, for particles annihilated in
vertex.

In the evaluation of $H_\lambda$, we have to calculate the effective
mass $m_{\lambda}$ for quarks and antiquarks.
Following~\citep{Glazek:1998sd}, we use the eigenvalue equation for a
single quark state, derived in 2nd order for $H_{\lambda}$, at one
arbitrary scale $\lambda = \lambda_0$ to specify notation for the
counterterm.
\begin{equation}
  \label{eq:m_tilde}
  \frac{P^{\perp\,2} + \tilde{m}^2}{P^+} \ket{q(P)}
  = \frac{P^{\perp\,2} + m^2_{\lambda_0}}{P^+} \ket{q(P)}
  - 
  H_{I\,\lambda_0}
  \frac{\ket{qg(P)}\bra{qg(P)}}{\mathcal{M}^2 - m^2}
  H_{I\,\lambda_0} 
  \ket{q(P)}.
\end{equation}
Here $\mathcal{M}^2$ is the free invariant mass of quark--gluon
states.  We express $m_{\lambda_0}$, presumably infinite when the
cutoff $\delta$ is removed, in terms of the would--be perturbative
eigenvalue $\tilde{m}^2$.  With hindsight, we set the eigenvalue mass
$\tilde{m}$ to be equal to the constituent quark mass.
Note that this condition is not meant to imply that there are free
quarks, since it is imposed only in the formal expansion in $g$, and
ignores nonperturbative effects that prevent quarks from being free in
this formulation.

We use $m^2_{\lambda_0}$ as the starting point in evaluation of
$m^2_{\lambda}$.  Obtained quark mass 
$m^2_{\lambda} = \tilde{m}^2 + \delta m^2_{\lambda}$ 
diverges logarithmically when $\delta$ goes to 0,
\begin{equation}
  \label{eq:dm_lambda}
  \delta m^2_{\lambda}
  \xrightarrow[\;\delta \to 0\;]{}
  \frac{g^2 C_F}{2 (2 \pi)^3}
  \sqrt{\frac{\pi}{2}}\lambda^2
  \log\frac{1}{\delta}.
\end{equation}
Here $C_F$ is the SU(3) color factor for color triplet.  We would
obtain the same result for $\delta m^2_{\lambda}$ in our approach if
we used the coupling coherence
condition~\citep{Perry:1997uv,Brisudova:1996hv,Perry:1993gp}.

Let us now consider the quark--antiquark interaction term in the
renormalized Hamiltonian $H_{\lambda}$.  For simplicity, we assume
that all quarks have the same masses but different flavors (to avoid
the insignificant discussion of antisymmetrization in the later
treatment of the van der Waals forces).  We will also use non
relativistic (NR) approximation, which much simplifies further
calculations.  This approximation is justified here when studying the
bound states of heavy quarks because the formfactor $f_{\lambda}$ with
$\lambda \ll \tilde{m}$ limits quark momenta to nonrelativistic values
despite a large coupling constant~\citep{Glazek:2002jg}.  We use a
three--vector $\vect{q}_{12}$ defined through the requirement that
\begin{equation}
  \label{eq:NR-notation}
  4 (\vect{q}_{12}^2 + m^2) = 
  \mathcal{M}^2_{12} = 
  \frac{\kappa^{\perp\,2} + m^2}{x(1-x)},\quad
  \vect{q}_{12}^{\perp} = \vect{\kappa}^{\perp}.
\end{equation}
From this requirement one obtains $x$ as function of $q_{12}^z$ and
$\vect{q}_{12}^{\perp}$, and one expands $q_{12}^z$ in the powers of
$\abs{\vect{q}_{12}}/m$, obtaining in first order
\begin{math}
  q_{12}^z \simeq \left(x - \frac{1}{2}\right) 2 m 
\end{math}
We should express the bare mass $m$ in terms of $\tilde{m}$, but
because the difference between $m$ and $\tilde{m}$ is of second order
in $g$ (see Eq.~\eqref{eq:dm_lambda}), and the interaction term is
already of second order, we can put $m = m_{\lambda} = \tilde{m}$
here.

When one applies the NR approximation to the effective potential
$f_{\lambda} V_{q\bar{q}}$ in Hamiltonian $H_{\lambda}$, one obtains
the following result (cf.~\citep{Brisudova:2001yc})
\begin{multline}
  \label{eq:V-interaction-q:NR}
  f_{\lambda} V_{\text{NR}}(\vect{q}) = 
  -g^2
  F_{12} \cdot F_{34}
  \exp
  \left[
    - \frac{\vect{q}^2_{12} - \vect{q}^2_{34}}{\lambda^4}
  \right]
  \times\mbox{} \\ \mbox{}\times
  \left\{
    \frac{1}{\vect{q}^2}
    + 
    \biggl[
      - \frac{1}{\vect{q}^2} + \frac{1}{q_z^2}
    \biggr]
    \exp\biggl(
      - \frac{2 m^2}{q_z^2}
      \frac{\bigl(\vect{q}^2\bigr)^2}{\lambda^4}
    \biggr)
  \right\}
  r^2_{\delta}\!\left(\frac{q_z}{2 m}\right),
\end{multline}
where $\vect{q} = \vect{q}_{12} - \vect{q}_{34}$.  The NR potential is
spin--diagonal.

The analysis of this interaction is made somewhat complicated by the
external similarity factor $f_{\lambda}$, which depends not only on
the momentum transfer \(\vect{q}\), but on $\abs{\vect{q}_{12}}$ and
$\abs{\vect{q}_{34}}$ separately.  However, when $\lambda$ is much
larger than momentum width of a wavefunction, $f_{\lambda}$ can be
replaced by 1 in the region that matters.

Following~\citep{Perry:1997uv,Brisudova:1996hv}, we neglect in our
analysis of van der Waals forces all Fock sectors except Q\=Q and
QQ\=Q\=Q.  This is justified for heavy quarks by the expectation that
gluons are lifted up in energy by certain gap condition, which is a
non-perturbative effect.  According
to~\citep{Perry:1997uv,Brisudova:1996hv}, gluons should not appear
explicitly in the resulting model.

When gluons are neglected, there remains uncanceled
$\delta$--divergent, $\lambda$--dependent term in effective quark
(antiquark) mass term.  This term cancels exactly in the matrix
element $\bra{\Psi_1} H \ket{\Psi_2}$ against the divergence coming
from Q\=Q--interaction, when $\ket{\Psi_1}$ and $\ket{\Psi_2}$ are
color singlets.  For other color states, the Hamiltonian matrix
elements are positive infinite, since the mass diverges stronger than
the potential of
Eq.~\eqref{eq:V-interaction-q:NR}~\citep{Perry:1997uv}.  Therefore,
the Hamiltonian is bounded from below for all states (as opposed to
phenomenological additive color potential model where 
$\bra{\Psi_1} H \ket{\Psi_2}$ goes to $-\infty$ with growing distance
between quarks  when both $\ket{\Psi_i}$ are in color--octet state,
for all \(V(r) \xrightarrow[r \to \infty]{} \infty\)~\citep{Greenberg:1981xn}).  
This is the first important distinction between the renormalization
group approach to QCD and potential models, where the colored states
would cause trouble.

Note however that this cancellation does not occur outside the NR
approximation~\citep{Brisudova:2001yc}.  One has to include
transitions to $q\bar{q}g$ sector to obtain cancellation outside NR
approximation.  The second order perturbation calculations would then
also lead to reduction of the confining part of effective potential,
but nonperturbative effects could prevent complete
cancellation~\citep{Glazek:privcomm}, resulting only in weaker force.

Color van der Waals forces between two mesons come from the mixing of
octet--octet state (coupled to overall singlet) to singlet--singlet
state, because the color dependent interaction polarizes mesons in
color space as well as in position space.  We choose states
$\ket{\alpha} = \ket{1_{13} 1_{24}}$ (both mesons in singlet state)
and $\ket{\beta} = \ket{1_{14} 1_{23}}$, as the basis of the color
subspace, where mesons are in color singlet state as a whole, where
$1_{ij}$ means that quark $i$ and antiquark $j$ are in color singlet.

We will use static treatment to see if small--$x$ divergences (which
are constants) cancel in the eigenequation for QQ\={Q}\={Q} sector.
For the meson--meson states, the interaction term $H_I$ is a $2 \times 2$ 
matrix in color space~\citep{Lipkin:1973pq}:
\begin{subequations}
  \label{eq:vdW:colors}
  \begin{align}
    H_I \ket{\alpha} &= 
    (\tfrac{8}{3} u_{\alpha} - \tfrac{1}{3}u_{\beta} + \tfrac{1}{3}u_{\mathrm{q}}) \ket{\alpha}
    + (u_{\beta} - u_{\mathrm{q}}) \ket{\beta}, \\
    H_I \ket{\beta}  &= 
    (u_{\alpha} - u_{\mathrm{q}}) \ket{\alpha}
    + (\tfrac{8}{3} u_{\beta} - \tfrac{1}{3}u_{\alpha} + \tfrac{1}{3}u_{\mathrm{q}}) \ket{\beta},
  \end{align}
\end{subequations}
where
\begin{align}
  \label{eq:u_something}
  u_{\alpha} &= u_{13} + u_{24}, &
  u_{\beta}  &= u_{14} + u_{23}, &
  u_{\mathrm{q}} &= u_{12} + u_{34},
\end{align}
and $u_{ij} = V(\vect{r}_{ij}) = V(\vect{r}_i - \vect{r}_j)$ is the
quark--(anti)quark potential.  The mass term is diagonal in
$\ket{\alpha}, \ket{\beta}$.  The divergences in the off diagonal
(mixing) part cancel out because $V_{\bar{q}\bar{q}} = V_{qq} = - V_{q\bar{q}}$.  
In the diagonal part, divergences in masses cancel
with divergences in potential inside singlets (i.e. $u_{\alpha}$ and
$u_{\beta}$, respectively) and divergences in the $u_{\mathrm{q}} - u_{\beta}$ 
and $u_{\mathrm{q}} - u_{\alpha}$ cancel also.  Thus, all
small--$x$ divergences cancel out and Eq.~\eqref{eq:vdW:colors} has
finite elements.  Therefore, color van der Waals forces emerge in the
approach presented here as they appear in color potential models.
This is the second result of the effective particle approach in the
approximation proposed in~\citep{Perry:1997uv,Brisudova:1996hv}.  The
third result is that we may then use the results on color van der
Waals forces as known from potential models.

To see what the van der Waals force looks like we have to look at the
NR potential~\eqref{eq:V-interaction-q:NR} behavior at large
distances.  We will use the spatial derivatives of the confining
potential~\citep{Perry:1997uv}.  The partial derivative with respect
to $r_x$ for $r_z = r_y = 0$, $\vect{r}$ being the distance between 
Q and \=Q in quarkonium state, is given by the equation
\begin{equation}
  \label{eq:V_conf:large-r_perp}
  \frac{\partial}{\partial r_x} V_{\text{conf}}(r_x)
  \simeq
  \frac{\alpha_s}{2 \pi}
  \frac{\lambda^2}{m}
  2 \sqrt{2 \pi}
  \frac{\partial}{\partial r_x}\log\abs{r_x/r_0}
  + \text{short range}
\end{equation}
(the result for the $r_y$ direction is identical).  Similar
calculations of the shape of the potential in the $r_z$ direction give
in the large distance limit the same functional form of the potential
but twice weaker.  Equation~\eqref{eq:V_conf:large-r_perp} contains
unknown value of $r_0$, but from the fit to numerically obtained
Fourier transformation one gets $r_0$ of the order of $m/\lambda^2$.

To simplify further calculations, we will use $V_{\text{conf}}$
averaged over angles (like in~\citep{Brisudova:1996hv}).  Numerical
calculations show that with growth of $r$ the range of polar angle
$\varTheta$, where $V_{\text{conf}}$ differs significantly from the
value for $\varTheta = \pi/2$, gets smaller and smaller.  Therefore,
for the large $r$ we can use the analytical result for the
$\vect{r}_{\perp}$ direction~\eqref{eq:V_conf:large-r_perp}, getting
the potential of the form
\begin{equation}
  \label{eq:V_conf}
  u(r) = V_0 \log(r/r_0), 
  \quad
  \text{where\ }V_0 = 2 \alpha_s\lambda^2\big/[\sqrt{2 \pi} m].
\end{equation}

For such logarithmic potential one obtains from the eigenvalues of $2
\times 2$ matrix in Eq.~\eqref{eq:vdW:colors} and virial
theorem~\citep{Greenberg:1981xn} the following color van der Waals
force between two heavy mesons:
\begin{equation}
  \label{eq:vdW:log}
  V_{\text{vdW}}(R) = 
  \frac{3}{54} V_0 \frac{\langle r^2_{13} \rangle^2}{{R}^{4}} \frac{1}{\log(R/r_{13})},
\end{equation}
where $R$ is the distance between the mesons and $r_{13}$ is the mean
width of a meson.  Note that the value of $r_0$ does not enter the
expression for the van der Waals force~\eqref{eq:vdW:log}.

The fit to the $B$ meson spectra via heavy quark effective theory
in~\citep{Brisudova:1996hv} and the fit to the $c\bar{c}$ and
$b\bar{c}$ spectra in~\citep{Brisudova:1997vw} give 
$\alpha_s \lambda^2/m$ around 1~GeV, i.e.\ $V_0 \simeq$~0.8~GeV.  
In color potential model~\eqref{eq:potential-model} with 
$V(r) = V_0 \log(r/r_0)$, one obtains from charmonium data 
$V_0 =$~731~GeV~\citep{Eichten:1980ms,Kwong:1987mj}.  
The question is then if such well established value of $V_0$ produces
acceptable strength of color van der Waals potential~\eqref{eq:vdW:log}.

We have done our calculations of the effective confining potential in
the NR limit, which strictly speaking can be valid only for heavy
quarkonia.  To the author's best knowledge there currently exist no
data on forces between such mesons.  Therefore, we will use existing
experimental limits on the color van der Waals forces between
\emph{light hadrons}, to make only order of magnitude guesses about
the allowed strength of $V_0$.

Between two (heavy) baryons one obtains the same type of force as
between two (heavy) mesons but with different coefficient (e.g.\ 
\citet{Gavela:1979zu} gives $1/2$ for baryons instead of $3/54$ for
mesons, but they use a different method to calculate van der Waals
force coming from potential~\eqref{eq:V_conf}).  Our NR
potential~\eqref{eq:V-interaction-q:NR} does not apply to light
quarks, but because van der Waals force is a low--energy interaction
acting between slow--moving particles, and in constituent models NR
approximation is commonly used, we will treat the NR
equation~\eqref{eq:vdW:log} for the van der Waals potential as
providing approximately right order of magnitude, disregarding any
doubts that such procedure may no longer be adequate in the case of
effective QCD for light quarks, where precise comparison with data is
attempted.

Data from Cavendish--type experiments~\citep{Feinberg:1979yw} give no
real constraints on the van der Waals force resulting from logarithmic
inter--quark potential, as opposed to the one from linear
potential~\citep{Greenberg:1981xn}.  The data from hadronic
atoms~\citep{Feinberg:1979yw} implies $(3/54) V_0 \lesssim 2\ \text{MeV}$ 
i.e.\ $V_0 \lesssim 40\ \text{MeV}$ for $r_{13} = 1\ \text{fm}$.  
We see that $V_0 \simeq 0.8\ \text{GeV}$ appears to be
very large.  \citet{Sawada:2000sm,Sawada:2000td} finds in the behavior
of P--wave amplitude of $\pi$--$\pi$ and S--wave p--p scattering at
low energy some signs of long range super--strong force $\propto
1/R^{n}$ with $n$ about 6, but no trace of forces $\propto 1/[R^4\log(R)]$ 
is found.  While the strength of his force is too large to
be explained using $\alpha_s$ that appears in Eq.~\eqref{eq:V_conf},
and remains to be explained, it is interesting that van der Waals
force $\propto 1/R^6$ comes from Coulombic part of the inter--quark
potential, which cannot be easily canceled.

It is clear that the calculation reported here can be considered only
an initial step towards derivation of color van der Waals forces in
QCD.  The non--cancellation of small--$x$ divergences without NR
approximation, and the lack of rotational symmetry of $V_{q\bar{q}}$,
suggest that one needs to consider Fock sectors with additional
gluons.  Most prominently however, one should expect corrections from
creation of quark--antiquark pairs~\citep{Brisudova:1998wq}.  These
elements can reduce the van der Waals forces we obtained here and
recover agreement with experimental findings.  The most attractive
feature of presented approach is that, contrary to the potential
models, the effective particle picture is clearly open to further
study in QCD, and can incorporate all these effects starting from
first principles.

\bibliography{vdW-bib}

\end{document}